# Light-controlled skyrmions and torons as reconfigurable particles

HAYLEY R. O. SOHN,[1] CHANGDA D. LIU,[1] YUHAN WANG,[1] AND IVAN I. SMALYUKH[1,2,3*]

[1]*Department of Physics and Materials Science and Engineering Program, University of Colorado, Boulder, CO 80309, USA*
[2]*Department of Electrical, Computer, and Energy Engineering and Soft Materials Research Center, University of Colorado, Boulder, CO 80309, USA*
[3]*Renewable and Sustainable Energy Institute, National Renewable Energy Laboratory and University of Colorado, Boulder, CO 80309, USA*
*[*ivan.smalyukh@colorado.edu](*ivan.smalyukh@colorado.edu)*

**Abstract:** Topological solitons, such as skyrmions, arise in field theories of systems ranging from Bose-Einstein condensates to optics, particle physics, and cosmology, but they are rarely accessible experimentally. Chiral nematic liquid crystals provide a platform to study skyrmions because of their natural tendency to form twisted structures arising from the lack of mirror symmetry at the molecular level. However, large-scale dynamic control and technological utility of skyrmions remain limited. Combining experiments and numerical modeling of chiral liquid crystals with optically controlled helical pitch, we demonstrate that low-intensity, unstructured light can control stability, dimensions, interactions, spatial patterning, self-assembly, and dynamics of these topological solitons.

## 1. Introduction

The dynamic behavior of self-reinforcing solitary optical wave packets called "solitons" has attracted a great deal of interest among physicists and mathematicians [1-8]. These non-spreading solitons maintain their spatially localized shape while propagating and typically emerge from a delicate balance of nonlinear and dispersive effects in the physical host medium, of interest for many applications, including telecommunications. Analogous phenomena involving solitons are also widely studied in fluid dynamics and many other physical systems [1]. Solitons of a different type, often called "topological solitons", are topologically nontrivial non-singular structures in continuous fields. They initially drew the interest of theorists in high energy physics and cosmology [9,10], but nowadays one can study them in light and condensed matter systems such as magnets, ferroelectrics, and liquid crystals (LCs) [11-23]. Being spatially-localized nonsingular field configurations [24-26], unlike the solitary waves mentioned above, topological solitons are not always associated with out-of-equilibrium dynamics, and instead are commonly also studied as static field configurations embedded in a uniform background or as topologically protected magnetic structures driven by electric currents in prototypes of magnetic racetrack memory devices [17,18]. In addition to many condensed matter and nuclear physics systems, a classical optical analog of a skyrmion-like structure has been recently reported in evanescent electromagnetic fields and demonstrated using surface plasmon polaritons [20]. Also very recently, photonic skyrmions in momentum space associated with equations of motions of a topological electromagnetic field have been studied theoretically [22], where the physics of this topological field configuration manifests itself in nontrivial windings of spin -1 vector as opposed to spin -1/2 vector [22]. In LCs topological solitons are associated with the corresponding spatial changes of the effective refractive index and optical axis orientation [27,28], determined by the orientation of the LC director, which makes them of interest for potential applications in diffractive optics, all-optical beam-steering, and a large variety of self-reconfigurable photonic devices. However, the means of controlling their location, translational dynamics, dimensions, and other properties and characteristics remain limited. In this report, we present an experimental approach for driving the dynamics and spatial patterning of topological solitons in confined chiral nematic LCs using patterned, unstructured light illumination by selectively manipulating the elastic free energy landscape within the host LC material to induce motion. Previous studies of LC skyrmions have outlined how to locally convert electric energy to motion at the scale of individual topological solitonic particles, giving rise to many different types of out-of-equilibrium behavior, including schooling of skyrmions [29-31]. However, this method requires electrodes on the sample surfaces in order to apply an oscillating electric field and has limitations in the controlled guiding of skyrmion motions: these motion directions are selected spontaneously, unless gradients of thickness or rubbing of homeotropic confining surfaces are used. Interestingly, we can also control the out-of-equilibrium behavior of these systems in a more hands-off manner in cells without electrodes by means of optical manipulation using patterned light illumination by employing specially selected photo-responsive chiral additives. This method of inducing optically powered skyrmion motions and robust control of their collective behavior is the focus of our present study. Since skyrmions can be created in light [20-23] and in material systems like liquid crystals and magnets [11-32], which, as we show, can be highly sensitive to light, our study may potentially lead to a possibility of exploring the interplay between the skyrmionic structures in light and in the condensed matter systems.

Our approach and findings may also lead to technological applications in diffractive optics, all-optical beam-steering, and self-reconfigurable photonic devices.

## 2. Topology and energetics of skyrmions and torons

Although both three-dimensional (3D) and two-dimensional (2D) analogs of Skyrme solitons have been realized in confined chiral nematics [11,19,25], here we will focus on the 2D case, often called the "baby skyrmion." These structures can be classified by a skyrmion number, defined as the integer number of times the vectorized director field configuration wraps the corresponding order parameter sphere, $\mathbb{S}^2$, which encompasses all possible directional unit vectors (Fig. 1a). The simplest type of baby skyrmion exhibits π-twist from the center to the periphery in all radial directions and, in mapping the field configuration onto the $\mathbb{S}^2$ sphere, covers it completely precisely once (Fig. 1a). Its skyrmion number is therefore one. We realize baby skyrmions of this type in chiral nematic LCs (Fig. 1), where the ground state has a helicoidal structure that is quantified by the cholesteric pitch, $p_0$, the distance over which the equilibrium director field **n(r)** of a ground-state chiral nematic LC twists by 2π. When we confine such a material between substrates with a cell-gap separation approximately equal to $p_0$ and perpendicular surface boundary conditions, the boundary conditions are incompatible with the one-dimensional helicoidal structure and that energetic frustration can lead to spontaneous formation of various twisted structures, including 2D skyrmions and torons [19,26]. In the case of finite-strength or weak boundary conditions in thin LC cells, degree-one 2D skyrmions typically form [19], with their structure (containing π-twist from the skyrmion's tube center to its periphery) being translationally invariant along the sample depth, with only minor perturbations close to the confining substrates [19]. In the case of strong perpendicular surface anchoring boundary conditions, the skyrmion tube terminates close to the confining substrates on two hyperbolic point defects, forming the elementary type of the so-called "toron" with an axisymmetric solitonic structure accompanied by a pair of self-compensating hyperbolic point defects (Fig. 1b) [26]. This elementary toron can be understood as a fragment of skyrmion terminating on point defects [29], similar to magnetic skyrmions that often terminate on Bloch points [33], and is the focus of our present study. LC skyrmions and torons have been analyzed extensively, including comparison of numerical simulations to experimental imaging, both with and without application of electric fields, to confirm understanding of the structures of such field configurations [19,29]. In this work, we extend this modeling to account for the facile response of the cholesteric pitch value to low-intensity optical illumination. The solitonic structures are modeled by minimizing the elastic free energy given by the Frank-Oseen expression,

$$W = \int \left\{ \frac{K_{11}}{2}(\nabla \cdot \mathbf{n})^2 + \frac{K_{22}}{2}\left[\mathbf{n}\cdot(\nabla\times\mathbf{n}) + \frac{2\pi}{p_0}\right]^2 + \frac{K_{33}}{2}\left[\mathbf{n}\times(\nabla\times\mathbf{n})\right]^2 - \frac{\varepsilon_0\Delta\varepsilon}{2}(\mathbf{E}\cdot\mathbf{n})^2 \right\} dV \quad (1)$$

where the Frank elastic constants $K_{11}$, $K_{22}$, and $K_{33}$ represent the elastic energy costs for splay, twist, and bend deformations of **n(r)**, respectively. $\Delta\varepsilon$ is the dielectric anisotropy and $\varepsilon_0$ is the permittivity of free space. By selectively increasing the pitch via blue-light excitation (Fig. 2) to a locally higher excited-state pitch value $p_e$ ($p_e > p_0$), the free-energy landscape within the exposed regions of the sample changes, with the free energy associated with embedding individual solitons increasing as compared to the sample areas that have not been illuminated. Because topological solitons represent minima in the free energy, they are sensitive to light and exhibit a broad range of facile responses, such as size changes and lateral translations, that are driven by free energy minimization and that we discuss below. We report on how to utilize the skyrmion's strong reaction to light exposure in order to control their self-assembly, dynamics, and spatial self-patterning behavior.

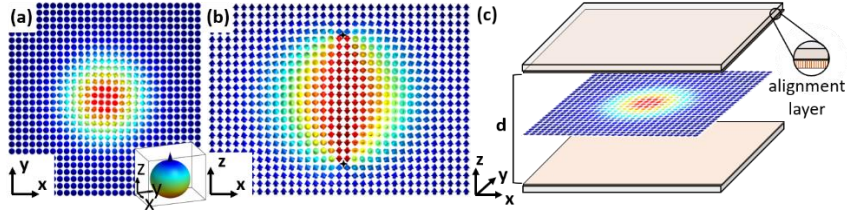

Fig. 1. Vectorized director field structure of the studied topological soliton. (a,b) Vectorized director structure of a skyrmion in the cross-sectional plane (a) orthogonal to the far-field director **n₀** and (b) parallel to **n₀**, where two hyperbolic point defects terminate the skyrmionic tube near the confining surfaces (black crosses), forming an elementary toron [19]. The structure was numerically modeled for a confined chiral nematic mixture (E7 – QL-76) with vertical surface boundary conditions. Arrows represent the vectorized director field, **n(r)**, colored according to corresponding points on the $\mathbb{S}^2$ sphere shown in the inset of (a). (c) Schematic representation of the experimental geometry, with cell thickness, d, labeled and the thin homeotropic alignment layer (see Methods) deposited on the confining glass substrates represented in orange (inset).

## 3. Materials & methods

Samples were prepared by mixing the QL-76 chiral additive (obtained from the Air Force Research Lab) [34] with an E7 nematic host (purchased from EM Industries), as the mixture is heated to the isotropic phase and agitated to ensure homogeneity. Various azo-chiral liquid crystalline materials have been previously synthesized and studied [35] and it is well-documented that an azobenzene-based chiral dopant can be reconfigured via a trans-cis isomerization upon UV- or blue-light excitation, leading to a change in the helical twisting power of the chiral dopant [35-37]. In our experiments, we chose to use an azobenzene-based chiral dopant QL-76 (Fig. 2a) [34,38]. The QL-76 additive, when mixed with a nematic host, exhibits a maximum helical twisting power decrease from a ground-state value of 60 µm$^{-1}$ to 27 µm$^{-1}$ upon excitation [38]. The ground-state helicoidal pitch, $p_0$, is set by controlling the concentration of the chiral dopant, $c$, and the helical twisting power, $h_{HTP}$, according to the relation $p_0 = 1/(h_{HTP} \times c)$ [39] (Table 1). Prior to cell construction, the surfaces of the glass slides are treated for perpendicular surface boundary conditions by dip-coating in a 1 wt % solution of DMOAP (Dimethyloctadecyl[3-(trimethoxysilyl)propyl]ammonium chloride, from Sigma Aldrich). The chiral mixture is then infiltrated into the glass cells, which have been glued together using small drops of UV-curable glue with dispersed glass spacers between two glass slides and curing for 60 s to set the cell gap constant (OmniCure UV lamp, Series 2000). Skyrmions were controllably generated in the E7 – QL-76 cells by means of optical reorientation using an optical laser tweezer setup comprised of a 1064 nm Ytterbium-doped fiber laser (YLR-10-1064, IPG Photonics) [11,12,19,25,26,29-31]. Large skyrmion lattices with and without lattice packing defects were produced by coupling the optical tweezers with a homemade LabView program in which lattice packing, spacing, and density can be pre-defined [27,40]. After generating individual skyrmions or multi-skyrmion arrays, we then use low-intensity (~ 1nW per square micrometer) [41,42] patterned blue light projection in the 425-480nm range to tune the pitch of an E7 – QL-76 mixture, in the range from ~10 to ⩽ 22 µm (Fig.2b,c).

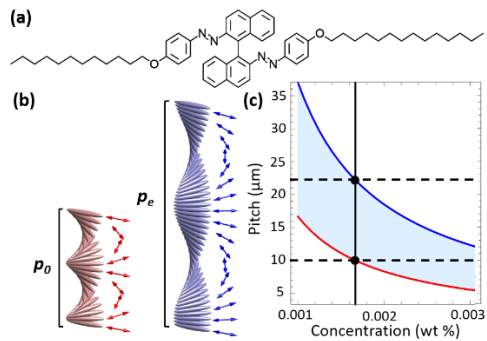

Fig. 2. Helicoidal structure of a photo-sensitive chiral nematic LC. (a) Molecular structure of the QL-76 chiral additive. [34] (b) Schematics of helicoidal structures for initial twisting rate defined by $p_0$ and maximum $p_e$ pitch value achieved upon illumination. (c) Pitch dependence on chiral additive (QL-76) concentration and photoexcitation, where the red line represents concentration-dependent $p_0$ and the blue line represents that of the maximum excited-state $p_e$. For the used E7 sample with QL-76, the black dashed lines mark the minimum and maximum possible pitch values that can be achieved at different light exposures.

Considering the facile response of photo-responsive chiral additives to blue-light illumination (Figs. 2 and 3) and the much weaker response to red light [38], we use spectrally-separated blue light to control solitonic structures and red light to image them at the same time and with a single integrated optical setup. Optical imaging was done using a BX51 Olympus upright microscope equipped with a charge-coupled device camera (purchased from Point Grey Research, Inc.), crossed polarizers above and below the sample, 4x, 10x, and 20x dry objectives (numerical apertures ranging from 0.3 to 0.9), and a red filter (Edmund Optics). Finely focused patterns of blue light were projected onto the samples using the LC micro-displays of an Epson EMP-730 LC Projector that were integrated with the BX51 microscope (Fig.4b) using lenses and a dichroic mirror (DM 505LP) that reflects the blue patterning light but transmits the red imaging light from the microscope light source [41]. With this setup, the blue patterning light is in the 450-480 nm range and has an intensity on the order of ~ 1 nW per square micron [41]. Because of the relatively long wavelength of the blue patterning light and the thin ~10μm cell thickness, absorbance within the LC sample doped with low concertation of additives is negligible and the illumination can be considered of constant intensity throughout the cell depth [38,43]. The experimental images and videos were analyzed using open source ImageJ/FIJI software (obtained from the National Institute of Health), through which positional data for each video frame was extracted using built-in particle tracking capabilities.

In order to obtain computer-simulated images of the director field configurations, we use a MATLAB-based numerical approach to minimize the Frank-Oseen free energy of the LC (Eqn. 1) [11,26,29]. Computer-simulated polarizing optical images are then based on these energy-minimizing structures with experimental material parameters, such as birefringence, $\Delta n$, and cell thickness, d, and generated using a Jones matrix method [29,44]. The material parameters used in all computer simulations correspond to independently characterized experimental values (Table 1), where $n_{ext}$ and $n_{ord}$ represent the extraordinary and ordinary refractive indices of the nematic host E7.

**Table 1: Material parameters for E7 nematic host and QL-76 chiral additive**.

| E7 Material Properties | |
|---|---|
| $\Delta\varepsilon$ | 13.8 |
| $h_{HTP}$ of QL-76 (μm$^{-1}$) {excited state} | 60 {27} |
| $K_{11}$ (pN) | 6.4 |
| $K_{22}$ (pN) | 3.0 |
| $K_{33}$ (pN) | 10.0 |
| $n_{ext}$ | 1.52 |
| $n_{ord}$ | 1.73 |
| $\Delta n$ | 0.21 |

## 4. Results and discussion

By selectively patterning μm$^2$-to-cm$^2$ areas within the chiral nematic LC samples using blue-light illumination, we gain optical control over the free-energy landscape within the chiral nematic sample. Because our skyrmions are stabilized by the chirality of the material, selective manipulation of the cholesteric pitch in various regions throughout the sample can tune their dimensions and dynamics. Therefore, when allowed to move, skyrmions and torons always tend to escape into the darker regions of illumination patterns, but responses can be even more complex if such dynamics is somehow hindered or if changes of pitch occur faster than these topological solitons can move. By harnessing these facile optical response effects, we demonstrate a high level of experimental control by probing individual skyrmions, large skyrmion lattices, and skyrmion bags, as detailed below.

*4.1 Skyrmion energetic adaptations to light illumination*

We investigate what happens to the skyrmions when the pitch is manipulated by full-cell illumination and increases consistently throughout the sample to neglect any gradient effects. We observe a clear linear dependence of the skyrmion's dimensions on the ratio d/p between the cell thickness, which remains fixed, and the pitch *p* that varies continuously from $p_0$ to $p_e$, depending on the exposure time. We systematically decrease d/p through blue light illumination (Fig. 3). By comparing computer-simulated structures (Fig. 3a-c) to experimental results (Fig. 3d,e) we can define a d/p stability range in which the skyrmions correspond to at least the local free energy minimum. Above d/p = 1.25 the skyrmion becomes more delocalized, with a stretched structure that is unfavorable for these experiments. Below d/p = 0.975 the skyrmion starts to shrink dramatically (Fig. 3) and eventually "pops" or collapses in on itself and disappears (Fig. 3d). We demonstrate the difference between a favorable and unfavorable sample based on the stability range (Fig. 3). For a small change in pitch upon 5s of blue light illumination, from $p_0 = 10$ µm to $p_e = 12.5$ µm, *Sample A* with thickness d = 10 µm has a starting d/p = 1 and an ending d/p = 0.8, which is outside of the stability range and the skyrmions therefore disappear (Fig. 3d). However, for the same 5s illumination and pitch increase from 10 µm to 12.5 µm, *Sample B* with thickness d = 12.5 µm maintains d/p within the stability range throughout, from d/p = 1.25 to 1, and the skyrmions not only remain stable, but also can be manipulated and pushed around with light without disappearing.

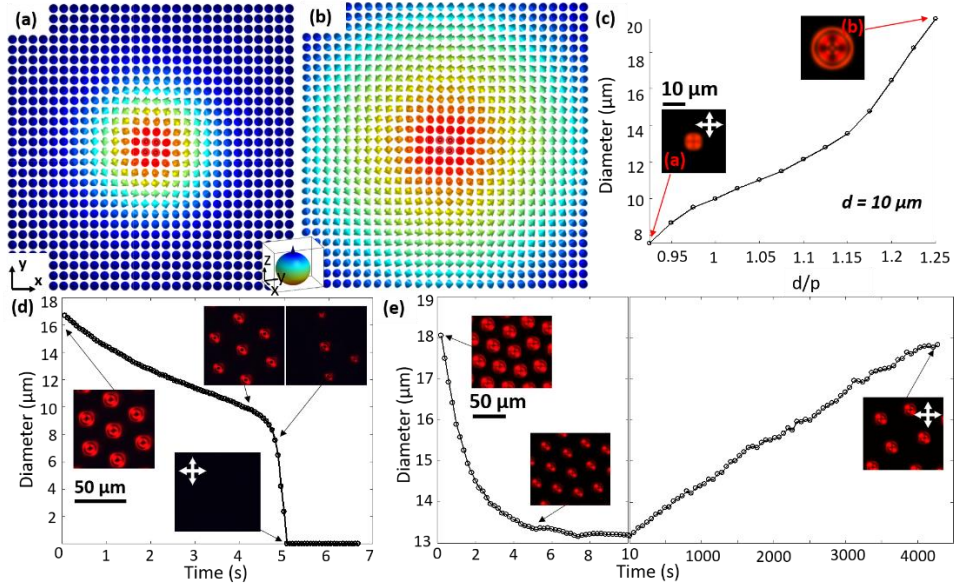

Fig. 3. Skyrmion dimensions with changing pitch. (a, b) Computer-simulated skyrmions in a chiral nematic LC shown at (a) d/p = 0.925 and (b) d/p = 1.25, where the vectorized **n(r)** is colored according to orientations on the $\mathbb{S}^2$ sphere (inset). (c) Simulated skyrmion diameter measured within the d/p stability range, where sample thickness is fixed at 10µm and corresponding diameters from (a) and (b) are marked in red, inset with simulated polarizing optical microscopy images corresponding to parts (a) and (b). (d, e) Experimentally measured skyrmion diameter upon 5s of blue-light exposure, with corresponding polarizing optical images (inset) shown for (d) *Sample A* and (e) *Sample B*. Crossed polarizers orientations are marked with white double arrows. Numerical simulations are based on material parameters of nematic host E7 and left-handed chiral additive QL-76 (see Methods). A similar E7 – QL-76 mixture was used in experiment, details of which are reported in Table 1.

Although the details of the director transformation during the transition from the skyrmion state to uniform unwound state are beyond the scope of the present study, it is worth mentioning that this transition invokes dynamics of singular point defects that eventually annihilate while "unzipping" the skyrmion tube in between them. Computer simulations reveal that this discontinuous transition from the skyrmionic structure to unwound homeotropic state occurs in the regime where the twisted skyrmionic state becomes more energetically costly than the unwound state (Fig. 4a). With this understanding of the energetic limitations of skyrmion manipulation, we choose to conduct experiments on samples with a slightly higher ground-state d/p, similar to *Sample B* described above, thus providing a larger stable parameter space for topology-preserving structural alteration of our skyrmions while avoiding the irreversible popping phenomenon.

Now that we understand the behavior and energetic limitations of skyrmions under full-area illumination, we expand our methods to probe skyrmion behavior under patterned illumination, in which we can create pitch- and energetic-gradients throughout the sample in the range shown in Fig. 4a, where lower stability ratios correspond to illuminated areas and higher ratios correspond to dark regions without illumination. With this patterned exposure (Fig. 4b), the skyrmions can "feel" (due to natural free energy minimization behavior) and adjust to the energetic gradient by moving towards a dark region with lower energy and lower pitch ($p_0$) to maintain their energetic stability, as demonstrated both numerically and experimentally (Fig. 4). The light-induced dynamics of a skyrmion can be carefully probed, leading to complex directional path motion over distances ~50-100x larger than the skyrmion's dimensions (Fig. 4d).

To verify that the light-powered motion does not inherently change the structure of the solitonic configurations, we turn again to computer simulations and view the skyrmions in the cross-sectional plane containing the far-field director (Fig. 5). These visualizations allow us to confirm that the entire localized structure remains the same, including the hyperbolic point defects near the substrates terminating the skyrmion tube of a toron, as it changes size at different d/p (Fig. 5a,b) and translates laterally through the sample (Fig. 5c,d). We therefore confirm that the structure of our LC skyrmions remains topologically unchanged under both full-area illumination and patterned illumination with the resulting motion.

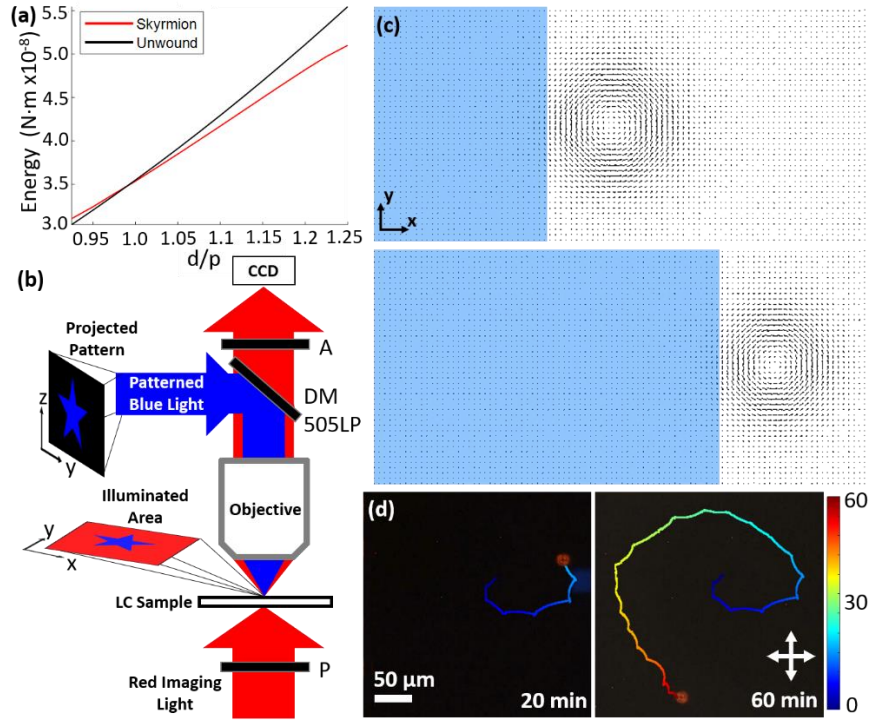

Fig. 4. Patterned light projection for controlled skyrmion motion. (a) Elastic free energy dependence of the twisted skyrmionic state (red) and the unwound or homeotropic state (black) on d/p. Energy was computed numerically on a 100μm x 100μm x 32μm computational volume. (b) Experimental setup for blue-light patterning projection with red imaging light, polarizer (*P*), analyzer (*A*), dichroic mirror (*DM 505LP*), and charge-coupled device camera (*CCD*, PointGrey, FlyCap) labeled. 4x, 10x, and 20x Olympus dry objectives were used. (c) Computer-simulated demonstration of skyrmion propagation, where the blue region represents illumination and black rods represent **n(r)** orientation. (d) Experimental patterned-light-induced motion of a skyrmion in a spiral path, the trajectory of which is shown overlaid on the polarizing images and colored according to elapsed time in seconds (right-side inset). Crossed polarizer and analyzer orientations are marked with white double arrows.

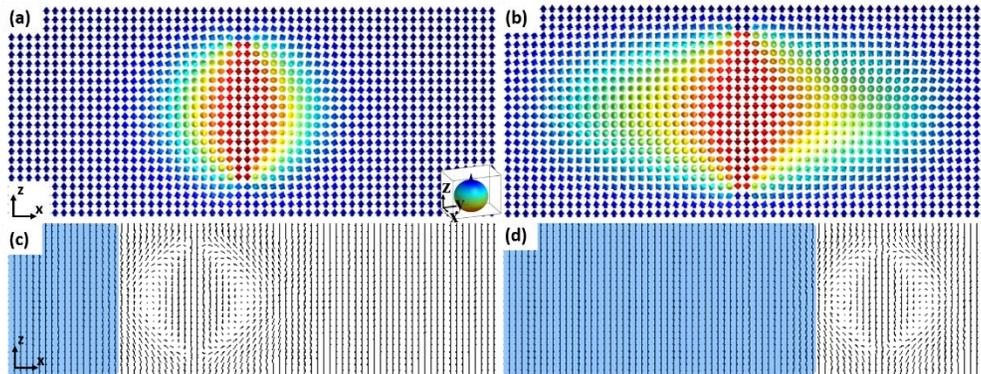

Fig. 5. Skyrmion-toron structure evolution with changing pitch. (a, b) Numerically simulated vertical cross-sections of torons in a chiral nematic LC shown in the cross-sectional plane containing the far-field director at (a) d/p = 0.925 and (b) d/p = 1.25, where the vectorized **n(r)** is colored according to orientation on the $\mathbb{S}^2$ sphere (inset). (c, d) Computer-simulated demonstration of photo-induced skyrmion propagation shown in the cross-sectional plane containing the far-field director, where the blue region represents the part of sample under illumination and black rods represent **n(r)** orientation. Numerical simulations are based on material parameters of nematic host E7 and left-handed chiral additive QL-76 (see Methods), with d = 10 μm. A similar E7 – QL-76 mixture was used in experiment, details of which are reported in Table 1.

*4.2 Control of symmetry and parameters of skyrmion lattices*

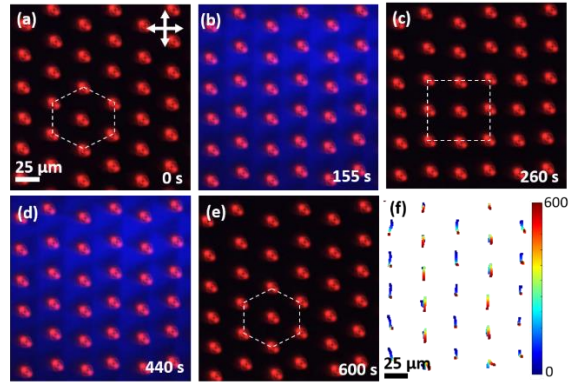

Fig. 6. Photo-induced skyrmion lattice rearrangement. (a-e) Polarizing optical microscopy images of (a) initial hexagonal skyrmion lattice before illumination, (b) a square-patterned blue-light illumination mask and (c) resulting square lattice skyrmion arrangement. (d) A hexagonal-patterned blue-light illumination mask and (e) resulting hexagonal lattice. (f) Trajectories of skyrmion motion throughout rearrangement, colored according to elapsed time in seconds (right-side inset). The time-coded motion trajectories are consistent with the motions of the dark regions in the illumination patterns. Crossed polarizer and analyzer orientations are marked with white double arrows and elapsed time is noted in the bottom right corners of images. The material is E7 doped with QL-76.

We demonstrate optically reconfigurable control over arrays of solitons and study an assortment of materials-science inspired demonstrations by generating various periodic lattices with large skyrmion densities (Fig. 6-8). By using a checkerboard-like illumination pattern we reversibly rearrange a hexagonal/triangular lattice of solitons (Fig. 6a) to a square-periodic lattice (Fig. 6d) and back (Fig. 6f) over the course of about 10 minutes. This is done by translating the illumination pattern with the skyrmions closely following the dark regions of the pattern. The direction of the pattern translation determines the direction of skyrmion motions because these solitons tend to stay in the dark regions of the pattern to keep their free energy at minimum. While skyrmions embedded in a uniform homeotropic background are known to interact repulsively [45], which is consistent with the hexagonal lattice under the conditions of a large number density of skyrmions per unit area [25], our findings show that light can drive assembly of out-of-equilibrium crystalline lattices of solitons, like the square-periodic lattice. Considering that skyrmions exhibit Brownian motions and particle-like behavior, the light-controlled solitonic assemblies can potentially provide a platform to investigate structural phase transitions in condensed matter, e.g. between equilibrium and out-of-equilibrium 2D crystalline lattices.

Interestingly, various types of defects can be embedded into these lattices as well. When we start from defining a grain boundary between two hexagonal lattices of topological solitons (Fig. 7) and expose the upper and lower boundaries, the energetic gradient becomes such that the inner part of the grain boundary is more energetically favorable for the skyrmions. The illumination therefore pushes the solitons together into a healed lattice without a grain boundary, towards an equilibrium state (Fig. 7a). Time-coded colored trajectories show how the solitons move smoothly towards each other (Fig. 7b) and how the many-body interactions between them eventually realize the lattice that corresponds to an equilibrium organization at large number-density of skyrmion particles. We also demonstrate precise control of the energetic gradient and resulting solitonic motion that allows for creation and healing of a small crack in a close-packed hexagonal lattice (Fig. 7c,d). The time-coded colored trajectories displayed in Fig. 7d once again show that we have extremely precise and tunable control over selected dynamics on the order of 10 µm, while the diameter of the solitons is only ~15 µm. Interestingly, only the skyrmions immediately surrounding the exposure pattern move away from the illumination, which is accompanied by their slight shrinking, while those at the periphery of the frame, farther away from the illumination, do not react significantly (Fig. 7d). Such local control of effective skyrmion dimensions within the periodic lattices may allow for detailed explorations of melting and crystallization transitions [46,47], both with and without defects.

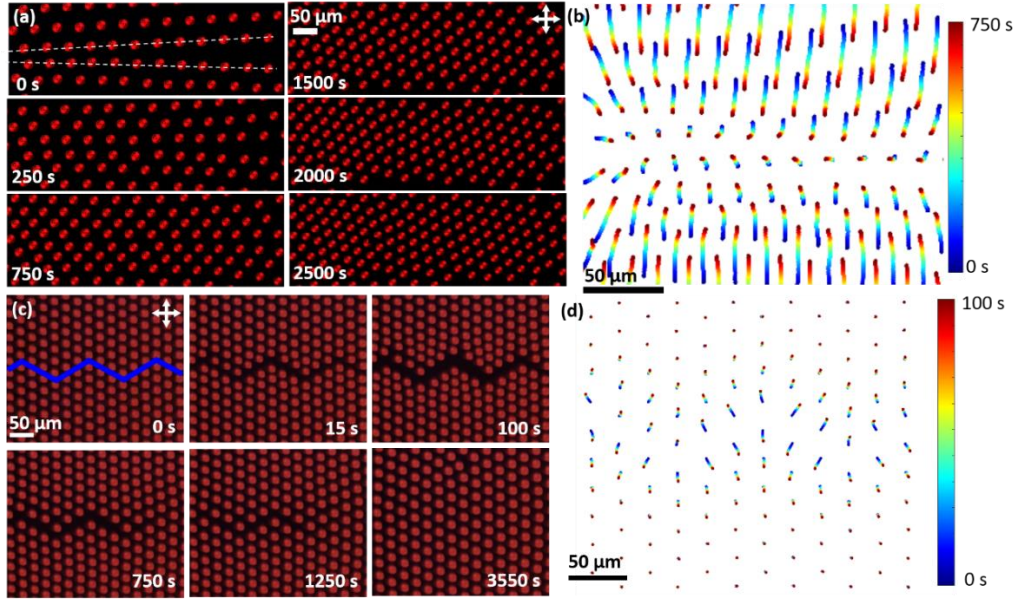

Fig. 7. Robust control of skyrmion lattices with spatially patterned light. (a) Polarizing optical images of a grain boundary in a skyrmion lattice and subsequent lattice healing upon illumination at the boundaries of the frame. White dashed lines mark the grain boundary in the initial frame. (b) Time-coded trajectories of skyrmion motion following illumination. (c) Polarizing optical images of formation of a photo-induced crack in a close-packed hexagonal lattice of skyrmions and subsequent lattice healing with time. The blue-light illumination pattern is shown as an overlay in the first frame. (d) Time-coded trajectories of the skyrmion manipulation to create a crack in the lattice. Crossed polarizer and analyzer orientations are marked with white double arrows and elapsed time is noted in the bottom right corners of images. The material is E7 doped with QL-76.

Additionally, we can induce compression-like transformations of skyrmion lattices by exposing the edges of the experimental frames (Fig. 8), for example starting from a square lattice and pushing the skyrmions together by creating an energetic gradient that favors the center of the frame and results in reorganization to a hexagonal lattice (Fig. 8a). Instead of a defect-free lattice, we can also start this compression process from a lattice with an edge dislocation pre-defined in the center (Fig. 8b). Illumination at the edges of the frame along both the vertical and horizontal axes induces compression that drives an interesting behavior that eventually results in a close-packed poly-domain hexagonal lattice (Fig. 8b). These demonstrations show that light could also be used to reconfigure optical solitonic gratings created out of LC skyrmions [27,28], which, in turn, could be used to control red or IR-range light, e.g. by creating optical vortices.

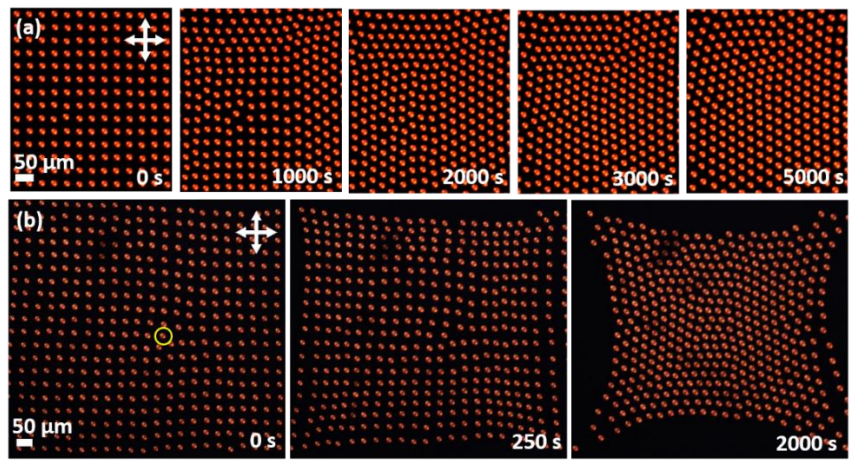

Fig. 8. Solitonic lattice compression with patterned light. (a, b) Polarizing optical images of (a) a square-periodic lattice being compressed by illumination at the boundaries of the frame and (b) a square lattice with an edge dislocation (marked with a yellow circle) being compressed on all sides by patterned illumination through a mask. Crossed polarizer and analyzer orientations are marked with white double arrows and elapsed time is noted in the bottom right corner of images. The material is E7 doped with QL-76.

### 4.3 Merging of skyrmion crystallites and skyrmion bags

To further explore photo-induced defect coarsening in skyrmion lattices, we use patterned illumination to drive nucleation of skyrmionic crystallites (Fig. 9). These pre-designed nucleation demonstrations were achieved by utilizing a large exposure pattern, wherein a circular region that was centered over the crystallites was dark, creating a radial energetic gradient that drives the skyrmions towards the center of the frame. As expected, the time scale of the packing defect coarsening is highly dependent on the initial orientation of the crystallites. Starting from crystallites that were drawn along the same lattice vector orientations (Fig. 9a), the defects coarsen soon after complete nucleation (Fig. 9b) with the skyrmions quickly adjusting to a hexagonal lattice due to repulsive interactions. However, when starting from crystallites that have been defined and optically oriented with a 20-degree offset in each of their lattice vectors (Fig. 9c), the coarsening of packing defects takes a much longer time (Fig. 9b) because the merging of the nucleated crystallites initially leads to the creation of grain boundaries. The evolution to an energetically favorable hexagonal-packed crystallite requires more rearrangement of skyrmions to align the crystallographic vectors by annealing the grain boundaries. The misalignment of LC or crystalline domains during ordered phase nucleation at the first-order phase transition is known to often lead to defects, with one example being the Kibble-Zurek mechanism of forming defects by merging nematic drops with misaligned director orientations [48,49]. Interestingly, although they are realized in chiral nematic hosts without intrinsic positional ordering, these skyrmionic particle-like structures may reveal details of how grain boundary defects form and behave during 2D crystallization transitions, mediating formation of polycrystalline materials.

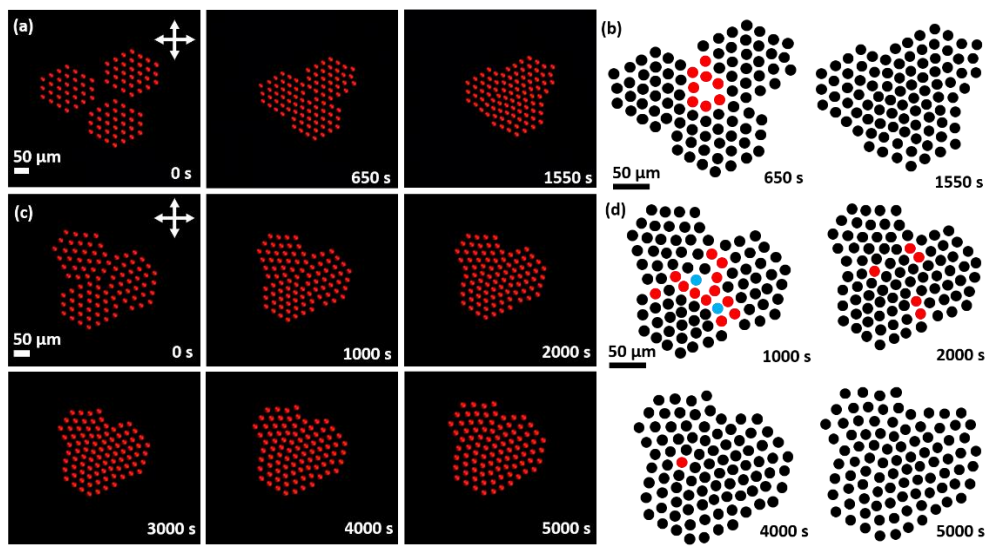

Fig. 9. Merging of skyrmion crystallites. (a) Polarizing optical microscopy images of three aligned skyrmion crystallites nucleating upon circular boundary illumination to form one large crystallite. (b) Schematics of aligned crystallites upon nucleation, with boundary skyrmions colored according to number of nearest neighbors (red = 5, black = 6). (c) Polarizing optical microscopy images of three misaligned skyrmion crystallites nucleating under similar illumination to form one large crystallite. (d) Schematics of misaligned crystallites upon nucleation, with boundary skyrmions colored according to number of nearest neighbors (blue= 4, red = 5, black = 6). Crossed polarizer orientation is marked with white double arrows and elapsed time is noted in the bottom right corner of images. The material is E7 doped with QL-76.

On the other hand, the interactions of elementary skyrmions in high energy and nuclear physics models allows for understanding the nature of subatomic particles with high baryon numbers [9,24]. While LC and magnetic 2D analogs of these skyrmions tend to mutually repel, it is possible to form high-degree condensed matter analogs of these high baryon number particles through formations of skyrmion bags and related structures [45]. This may again allow for using our soft matter system as a testbed for exploring and understanding complex topological phenomena. The skyrmion number of a skyrmion bag depends on the number of anti-skyrmions $(N_A)$ embedded within it, where a bag described by $S(N_A)$ has a topological degree $N_A - 1$ [45]. Here, using an approach similar to that described previously [45], we generate large skyrmion bags with skyrmion numbers close to 100 in our photo-sensitive cells as a means of studying their manipulation in the same way that we've characterized degree-one individual skyrmion photosensitivity. First, we probe the bag under full-area blue light illumination, similar to the experiments shown in Fig. 3. As the pitch increases across the entire sample area in the absence of an energy gradient, the skyrmions within the bag shrink in diameter, as expected, and the bag shrinks in size (Fig.10a). Similar to an individual skyrmion's behavior under these conditions, the bag grows back to it equilibrium dimensions with elapsed time in the absence of blue-light illumination. Next, we induce an energetic gradient by means of illumination on the left side of the frame with the goal of inducing controlled path motion of the skyrmion bag (Fig.10b). The resulting dynamic behavior is interesting and somewhat unexpected, owing to the internal pressure that the bag exerts on its inner skyrmions as it translates a short distance within the sample. As the pressure builds, with the left side of the bag favoring motion due to the energy gradient, the inner skyrmions begin to pop and eventually the bag shrinks down to an individual skyrmion, created by the bag collapsing in on itself (Fig.10b), again through a discontinuous process invoking singular defects that we discussed above. This demonstration shows us that high-degree skyrmionic topological structures react differently from individual baby skyrmions and can also be controlled by low-intensity illumination, including the control of the skyrmion number in a broad range of values within 1-100. This opens up many possibilities for the expansion and development of novel light-controlled topological materials in future works.

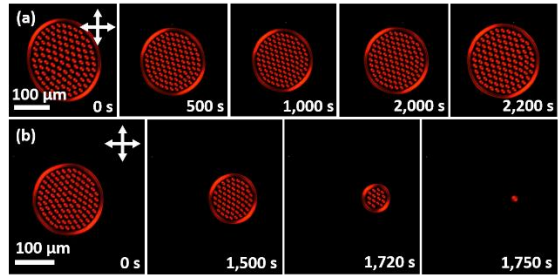

Fig. 10. Skyrmion bag manipulation and transformation. (a) An *S(102)* skyrmion bag shrinks following 5s of blue-light illumination, to the point of greatest-possible stable compression (where a single skyrmion pops through a transformation invoking singular defects), then grows back upon dark relaxation to a stable *S(101)* bag. (b) Upon blue light illumination on the left side of the frame, an *S(98)* bag starts to move slowly to the right, but the internal pressure of large-scale movement on the inner skyrmions causes soliton popping and over time the bag shrinks down to a single skyrmion. Crossed polarizer and analyzer orientations are marked with white double arrows and elapsed time is noted in the bottom right corners of images. The material is E7 doped with QL-76.

## 5. Conclusions

We have demonstrated optical control of particle-like topological solitons by low-intensity unstructured light, which can be used as a model system. Upon UV- or blue-light exposure, we observed two distinct responses of the skyrmionic structures. If the entire sample is exposed, the skyrmions do not feel any energetic landscape gradient and adapt their structures by shrinking in diameter and in some cases disappearing. However, when an illumination pattern is used to induce an energetic gradient in the sample, the skyrmions sense the free energy landscape and move towards dark regions of the cell without illumination, where the structures have lower elastic free energy. This twofold control allows us to implement many interesting materials-science and nuclear-physics inspired demonstrations, such as lattice rearrangement, grain boundary healing and crack propagation, lattice compression, crystallite merging during nucleation, and transformations of skyrmion bags. Considering that these skyrmions are hosted in anisotropic LC materials, our findings may lead to photo-responsive photonic gratings and diffraction patterns, privacy windows (note that the intensity of light utilized to control skyrmions is comparable to that of ambient light), and even new touch-screen technologies. Although various types of optical control of skyrmionics structures in condensed matter have been demonstrated previously by using specially designed beams of light [19,26,32,50,51], our work enables such control when using unstructured light of intensity comparable to that of ambient light, which can expand the potential technological uses of skyrmions in LCs.

## Funding

This research was supported by the National Science Foundation (NSF) through Grants No. DMR-1810513 368 (research) and No. DGE-1144083 (graduate research fellowship to H.R.O.S.) and ACI-1532235 and ACI-1532236 (computational user facilities at the University of Colorado Boulder).

## Acknowledgements

We thank T. Bunning and T. White for providing the QL-76 chiral dopant used throughout this study. We also thank P. Ackerman, C. Bowman, J-S. B. Tai, T. White, R. Voinescu, and Y. Yuan for useful discussions. I.I.S. acknowledges hospitality of the Newton institute at Cambridge University during his stay, when part of the manuscript preparation was completed.